\documentclass[conference]{IEEEtran}
\IEEEoverridecommandlockouts
\usepackage{cite}
\usepackage{amsmath,amssymb,amsfonts}\usepackage[caption=false,font=footnotesize,labelfont=rm,textfont=rm]{subfig}
\usepackage{algorithm}
\usepackage{algorithmic}
\usepackage{xhfill}
\usepackage{graphicx}
\usepackage{textcomp}
\usepackage{xcolor}
\usepackage{colortbl}
\usepackage{array}
\usepackage{amsthm}
\usepackage{stfloats}
\usepackage{adjustbox}
\usepackage{makecell}
\usepackage[colorlinks,linkcolor=blue,anchorcolor=blue,citecolor=blue,urlcolor=blue]{hyperref}

\usepackage[margin=0.58in,top=0.645in,bottom=0.985in,columnsep=0.21in]{geometry}

\newtheorem{lemma}{Lemma}

\newtheorem{proposition}{Proposition}

\makeatletter
\renewcommand{\maketag@@@}[1]{\hbox{\m@th\normalsize\normalfont#1}}%
\makeatother

\begin{document}

\title{\fontsize{23.3pt}{30pt}\selectfont Multi-Target Position Error Bound and Power Allocation Scheme for Cell-Free mMIMO-OTFS ISAC Systems
\thanks{This work was supported by the National Natural Science Foundation of China under Grant 62271167. (\emph{Corresponding author: Shaochuan Wu.})}
}

\author{
\IEEEauthorblockN{Yifei Fan, Shaochuan Wu, Haojie Wang, Mingjun Sun, and Jianhe Wang}
\IEEEauthorblockA{School of Electronics and Information Engineering, Harbin Institute of Technology, Harbin, China}
\IEEEauthorblockA{Email: yifan@stu.hit.edu.cn, scwu@hit.edu.cn, \{22B905021, 24B305036, 23S105108\}@stu.hit.edu.cn}

}

\maketitle

\begin{abstract}
This paper investigates multi-target position estimation in cell-free massive multiple-input multiple-output (CF mMIMO) architectures, where orthogonal time frequency and space (OTFS) is used as an integrated sensing and communication (ISAC) signal. Closed-form expressions for the Cram\'{e}r-Rao lower bound and the positioning error bound (PEB) in multi-target position estimation are derived, providing quantitative evaluations of sensing performance. To enhance the overall performance of the ISAC system, a power allocation algorithm is developed to maximize the minimum user communication signal-to-interference-plus-noise ratio while ensuring a specified sensing PEB requirement. The results validate the proposed PEB expression and its approximation, clearly illustrating the coordination gain enabled by ISAC. Further, the superiority of using the multi-static CF mMIMO architecture over traditional cellular ISAC is demonstrated, and the advantages of OTFS signals in high-mobility scenarios are highlighted.
\end{abstract}


\section{Introduction}
Integrated sensing and communication (ISAC) has emerged as a key enabling technology in the forthcoming 6G era, offering enhanced communication and sensing (C\&S) performance
while reducing hardware costs and alleviating spectrum congestion~\cite{Lu2024Integrated}. The two functionalities use shared resources and are co-designed in an ISAC system for coordination gains, representing a more profound integration paradigm~\cite{liu2022integrated}. Under this vision, the existing cellular networks are expected to be equipped with ubiquitous perceptive capability, evolving into perceptive mobile networks~\cite{zhang2021enabling}.

In an ISAC system based on cellular networks, the transmitter and receiver are typically co-located and both functions are performed by cellular access points (APs), characterizing a mono-static sensing configuration~\cite{Dehkordi2023Beam}. However, the single observation angle generated by the cellular AP can be easily blocked in complex propagation environments~\cite{Liu2024Cooperative}. In addition, the cellular networks often suffer from a fairness problem at cell edges, resulting in unreliable C\&S services for ultra-reliable applications, such as vehicle-to-everything (V2X) technology~\cite{elfiatoure2023cell}. To address the limitations inherent in cellular networks, cell-free massive multiple-input multiple-output (CF mMIMO) has emerged as a promising solution. As a representative of multi-static sensing, the CF mMIMO architecture enables multi-angle observations and achieves spatial diversity by leveraging geographically distributed ISAC transmitters and receivers~\cite{behdad2024multi}. Moreover, the cell edges can be completely eliminated in this architecture by ensuring uniform and ubiquitous service~\cite{ammar2021user}.

Given the sensitivity of the traditional orthogonal frequency division multiplexing (OFDM) signal to Doppler shifts in high-mobility applications such as V2X, the emerging orthogonal time frequency and space (OTFS) signal stands out as a superior candidate for the CF-ISAC systems. By modulating information symbols in the delay-Doppler (DD) domain, the OTFS signal exhibits robustness against delay and Doppler spreads~\cite{Gong2023Simultaneous}.

\subsection{Related Work}
The employment of the OTFS signal in CF-ISAC systems has been recently investigated in~\cite{Fan2024Power,Singh2025Target}. Particularly, the authors in~\cite{Fan2024Power} derived a closed-form spectral efficiency (SE) expression regarding optional sensing beams. A power allocation strategy was proposed to maximize the minimum communication signal-to-interference-plus-noise ratio (SINR) between users while a specified sensing SINR value was guaranteed. Further, in~\cite{Singh2025Target}, target detection performance was evaluated in a sensing-centric approach, where transmit power was optimized to maximize the sensing signal-to-noise ratio (SNR) while ensuring a required quality-of-service (QoS) for communication users.

\subsection{Contributions}
Sensing tasks mainly involve \emph{target detection} and \emph{parameter estimation}~\cite{Lu2024Integrated}. The aforementioned studies have focused on \emph{target detection} by constraining or maximizing the sensing SINR, while leaving the latter task — \emph{parameter estimation} unexplored. To fill this gap, this paper investigates multi-target position estimation, using the position error bound (PEB) as a sensing performance metric. To the best of the authors’ knowledge, the \emph{parameter estimation} performance of targets in the CF
mMIMO-OTFS ISAC systems has not been explored in the existing literature. The main contributions of this paper are summarized as follows.
\begin{itemize}
  \item[$\bullet$] By exploiting the coordination gain achieved by ISAC, closed-form Cram\'{e}r-Rao lower bound (CRLB) and PEB expressions for multi-target position estimation are derived. To facilitate efficient analysis and optimization of the position estimation performance, this study introduces a low-complexity PEB approximation, which is applicable to all OTFS-signal-based multi-static sensing systems.
  \item[$\bullet$] A max-min fairness problem is formulated to improve the communication SINR between users while satisfying a particular sensing PEB constraint and per-AP power constraints. To solve this non-convex problem, a novel power allocation algorithm is developed based on iterative convex optimization.
  \item[$\bullet$] The results validate the proposed PEB expression and its approximation, offering an intuitive illustration of the coordination gain of ISAC in sensing. In addition, the superiority of the CF architecture with high-density APs is highlighted by a comparison with the cellular ISAC.
\end{itemize}

$\emph{Notation:}$ Lowercase letters, boldface lowercase letters, and boldface uppercase letters denote scalars, column vectors, and matrices, respectively. The superscripts $\left( \cdot \right)^*$, $\left( \cdot \right)^{\mathrm{T}}$, $\left( \cdot \right)^{-1}$, and $\left( \cdot \right)^{\dagger}$ represent the conjugate, transpose, inverse, and conjugate-transpose operations, respectively. The operators $\mathrm{Tr\left( \cdot \right)}$, $\mathbb{E}\left\{ \cdot \right\}$, and $\odot$ denote the trace, expectation, and Hadamard product, respectively; $\left\lceil\cdot\right\rceil$ is the ceiling function, and $\operatorname{diag}\{\cdot\}$ returns a diagonal matrix. Finally, $\left\| \cdot  \right\|$ and $\left| \cdot \right|$ represent the vector and scalar Euclidean norms, respectively.

\section{System Model}
\begin{figure}[!t]
  \centering
  \includegraphics[width=2.6in]{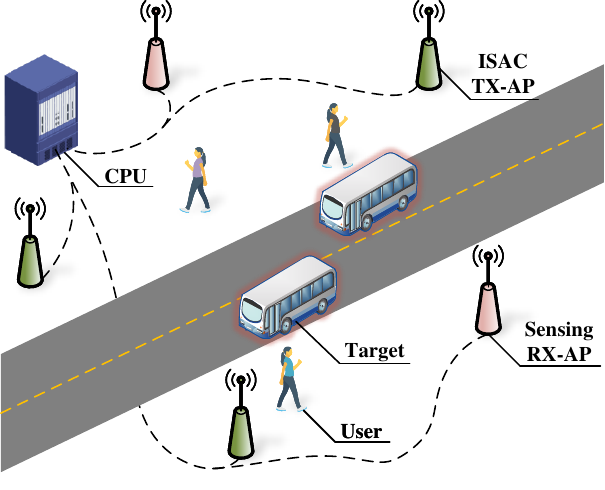}
  \vspace{-3mm}
  \caption{Illustration of the multi-target CF-ISAC system setup.}
  \label{fig:fig_1}
  \vspace{-12.5pt}
\end{figure}

This study considers a multi-target CF-ISAC system during the downlink phase, where the OTFS is used as an integrated signal. All $N_{\mathrm{AP}}$ APs are connected to a centralized processing unit (CPU) synchronously, and each AP is equipped with a uniform linear array (ULA) of $M_{\mathrm t}$ antennas. As depicted in Fig.~\ref{fig:fig_1}, each AP functions either as an ISAC transmitter or a sensing receiver, determined by a designed mode selection scheme. The $N_{\mathrm{tx}}$ transmitting APs employ maximum-ratio (MR) precoding to transmit integrated signals, jointly serving $K_{\mathrm u}$ single-antenna users while sensing $T_{\mathrm{g}}$ targets. The remaining $N_{\mathrm{rx}}$ receiving APs then collect the echo signals to estimate the targets' position.

\subsection{Downlink Communication Model}
The OTFS signal is assumed to have $M$ subcarriers with a subcarrier spacing of $\Delta f$, and $N$ symbols with a symbol duration of $T$. A cyclic prefix (CP) of sample length $N_{\mathrm{cp}}$ is added to each phase, ensuring the corresponding CP duration satisfies $T_{\mathrm{cp}}\geq\tau_{\max}$. The information symbols for the $q$th user ${x_q[k,\ell]}$ are scheduled on the DD grid $\Gamma=\big\{\frac{k}{NT},\frac{\ell}{M\Delta f}\big\}$. After performing the inverse symplectic finite Fourier transform (ISFFT) operation, the DD domain symbols $x_q[k,\ell]$ are converted to time-frequency (TF) domain as follows:
\vspace{-1mm}
\begin{equation} X_{q}[n,m] = \frac {1}{\sqrt {MN}}\sum _{k=0}^{N-1}\sum _{\ell =0}^{M-1} x_{q}[k,\ell] e^{j2\pi \left ({\frac {nk}{N}-\frac {m\ell }{M}}\right)},\end{equation} 
\vspace{-1mm}
\vspace{-1mm}

\addtocounter{equation}{7}
\begin{figure*}[hb] 
\vspace{-3mm}
\xhrulefill {black}{1pt}
  \centering
    \begin{small}
    \begin{equation}
        \Psi_{k,k^{\prime},l,l^{\prime}}\approx\frac{1}{NM}\!\sum_{n^{\prime}=0}^{N-1}\underbrace{\vphantom{\begin{array}{ll}1&l\in\mathcal{L}_{\mathrm{ICI}}(\tau):=[0,M-l_{\tau}-1]\\\\e^{-j2\pi\left(\nu T+\frac{k}{N}\right)}&l\in\mathcal{L}_{\mathrm{ISI}}(\tau):=[M-l_{\tau},M-1]\end{array}}e^{j2\pi(k^{\prime}-k+\nu NT)\frac{n^{\prime}}{N}}}_{\triangleq\alpha_{n^{\prime},k,k^{\prime}}(\nu)}
        \sum_{m^{\prime}=0}^{M-1}\underbrace{e^{j2\pi(l^{\prime}-l+\tau M\Delta f)\frac{m^{\prime}}{M}}e^{j2\pi\nu\frac{l^{\prime}}{M\Delta f}}\left\{\begin{array}{ll}1&l^{\prime}\in\mathcal{L}_{\mathrm{ICI}}(\tau):=[0,M-l_{\tau}-1]\\\\e^{-j2\pi\left(\nu T+\frac{k^{\prime}}{N}\right)}&l^{\prime}\in\mathcal{L}_{\mathrm{ISI}}(\tau):=[M-l_{\tau},M-1]\end{array}\right.\!\!\!}_{\triangleq\beta_{m^{\prime},k^{\prime},l,l^{\prime}}(\nu,\tau)}
        \label{eq:Psi}
    \end{equation}
    \end{small}
\end{figure*}
\addtocounter{equation}{-8}

\noindent for $n,k=0,\ldots,N-1$ and $m,l=0,\ldots,M-1$. Further, by performing the Heisenberg transform, $X_{q}[n,m]$ are converted to a time-domain signal as
\vspace{-1mm}
\begingroup\makeatletter\def\f@size{9.5}\check@mathfonts
\begin{align} 
\hspace {-0.4pc}s_{pq}(t)=\sum_{n=0}^{N-1}{\sum_{m=0}^{M-1}}\sqrt{\eta _{pq}}{X_q}[n,m]g_{tx}(t-nT)e^{j2\pi m\Delta f(t-nT)},
\label{eq:s_t}
\end{align}
\endgroup
\vspace{-1mm}
\vspace{-1mm}
\vspace{-1mm}

\noindent where $g_{tx}(t)$ is the transmitting pulse-shaping filter, $\eta _{pq},\, p=1,\ldots,N_{\mathrm{tx}},\,q=1,\ldots,K_{\mathrm{u}}$ are the power control coefficients set to make the average transmit power $P_p$ at each transmitting AP satisfy the following power constraint 
\begingroup\makeatletter\def\f@size{9.5}\check@mathfonts
\begin{equation} 
\vspace{-1mm}
P_p=\mathbb {E}\left \{{\Bigg| \sum _{q=1}^{K_{\mathrm{u}}}s_{pq}(t)+\sum _{t=1}^{T_{\mathrm{g}}}s_{pt}(t)\Bigg|^{2}}\right \} \leq P_{\mathrm{d}}.\label{eq:power_constraint}\end{equation}
\endgroup
\vspace{-1mm}

\noindent where $s_{pt}(t)$ is the time-domain sensing signal for target $t$ obtained by applying a similar procedure as in~\eqref{eq:s_t}, and $P_{\mathrm{d}}$ denotes the maximum downlink transmit power.

Considering the doubly selective fading caused by high user mobility, the DD domain channel impulse response from transmitting AP $p$ to user $q$ can be expressed as~\cite{Fan2024Power}
\vspace{-1mm}
\begin{equation} \mathbf{h}_{pq}(\tau,\nu) = \sum _{i=1}^{ L_{pq}} \mathbf{h}_{pq,i}\delta (\tau - \tau _{pq,i})\delta (\nu - \nu _{pq,i}),\label{eq:h_pq}\end{equation}
\vspace{-1mm}
\vspace{-1mm}

\noindent where the channel vector $\mathbf{h}_{pq,i}\sim \mathcal{CN}(\mathbf{0},\mathbf{R}_{pq,i})$ follows a correlated Rayleigh fading model, with its spatial correlation matrix $\mathbf{R}_{pq,i}=\mathbb{E}\{\mathbf{h}_{pq,i}^{\phantom{\dagger}}\mathbf{h}_{pq,i}^{\dagger}\}\in \mathbb{C}^{M_{\mathrm t}\times M_{\mathrm t}}\vphantom{\underline{\overline{\hat{\mathbf{R}}^{\dagger}_{pq,i}}}}$ reflecting the effect of geometric path loss~\cite{behdad2024multi}. Moreover, $\tau_{pq,i}$, $\nu_{pq,i}$, and $L_{pq}$ represent the $i$th path's delay, Doppler shift, and the number of paths from transmitting AP $p$ to user $q$, respectively.




From the downlink communication perspective, the transmitting APs apply MR precoding to transmit integrated signals to serve $K_{\mathrm{u}}$ users. The signal received at the $q$th user is given by
\vspace{-1mm}
\begin{align}
r_{q}(t) &=\nonumber \int_\tau \int_\nu \sum_{p=1}^{N_{\mathrm{tx}}} \mathbf{h}_{pq}^\mathrm{T}(\tau,\nu)\Bigg( \sum _{q^\prime=1}^{K_{\mathrm{u}}}\hat{\mathbf{h}}_{pq^\prime}^{*}(\tau,\nu)s_{pq^\prime}(t-\tau) \\
&\!\!+\!\sum _{t=1}^{T_{\mathrm{g}}}\hat{\mathbf{h}}_{pt}^{*}(\tau,\nu)s_{pt}(t-\tau) \!\Bigg)e^{j2\pi \nu (t-\tau)} d\tau d\nu + w_{q}(t),\!
\end{align}
\vspace{-1mm}
\vspace{-1mm}

\noindent where $\hat{\mathbf{h}}_{pq}$ is a unit-norm estimate of a channel vector $\mathbf{h}_{pq}$, $\hat{\mathbf{h}}_{pt}$ is the sensing precoding vector given by~\eqref{eq:sensing_BF} in the next section, and $w_{q}(t)\sim\mathcal{CN}(0,\sigma_w^2)$ is the noise received by the $q$th user. After performing the Wigner transform equipped with a receiving filter $g_{rx}(t)$, the TF domain received samples are obtained by a sampler as follows:
\vspace{-1mm}
\begin{equation} 
Y_{q}[n,m]=\int r_{q}(t) g_{rx}(t-nT) e^{-j2\pi m \Delta f (t-nT)} dt.
\end{equation}
\vspace{-1mm}
\vspace{-1mm}
\vspace{-1mm}
\vspace{-1mm}

{\noindent Finally, by applying the SFFT to $Y_{q}[n,m]$ and assuming non-ideal rectangular windows are used in the transmitting and receiving pulse-shaping filters, the DD domain signal received\parfillskip=0pt\par}

\noindent at the $q$th user can be formulated in a vector form as
\vspace{-1mm}
\begingroup\makeatletter\def\f@size{9.5}\check@mathfonts
\begin{align} 
\mathbf{y}_{q}=&\underbrace{\sum _{p=1}^{N_{\mathrm{tx}}} \eta _{pq}^{\frac{1}{2}} {\mathbf{H}}_{pq} \hat {\mathbf{H}}_{pq}^{\dagger} {\mathbf{x}} _{q}}_{\text{Desired signal}} \nonumber + \underbrace{\sum _{p=1}^{N_{\mathrm{tx}}} \sum _{q'\neq q}^{K_{\mathrm{u}}} \eta _{pq'}^{\frac{1}{2}}{\mathbf{H}}_{pq}\hat {\mathbf{H}}_{pq'}^{\dagger} {\mathbf{x}} _{q'}}_{\text{Inter-user interference}} \\[-4pt]
&+\, \underbrace{\sum _{p=1}^{N_{\mathrm{tx}}} \sum _{t=1}^{T_{\mathrm{g}}} \eta _{pt}^{\frac{1}{2}}{\mathbf{H}}_{pq}\hat {\mathbf{H}}_{pt}^{\dagger} {\mathbf{x}} _{t}}_{\text{Sensing interference}} + \underbrace{\vphantom{\sum _{p=1}^{N_{\mathrm{tx}}}}{\mathbf{w}}_{q}}_{\text{Noise}},\label{eq:y_ps}\end{align}
\vspace{-1mm}


\noindent where ${\mathbf{w}}_{q} \sim \mathcal{C N}\left(\mathbf{0}_{MN}, \sigma_w^2 \mathbf{I}_{MN}\right)$ \endgroup is the noise vector at user $q$, and the effective DD domain channel between the $p$th transmitting AP and the $q$th user is given by~\cite{Dehkordi2023Beam}
\vspace{-1.5mm}
\begingroup\makeatletter\def\f@size{9.5}\check@mathfonts
\begin{equation} {\mathbf{H}}_{pq} = \sum _{i=1}^{ L_{pq}} \left( \mathbf h_{pq,i}^\mathrm{T}\otimes {\boldsymbol{\Psi}} _{pq}^{(i)} \right)\in \mathbb{C}^{MN\times M_{\mathrm t}MN},\label{eq:DD_channel}\end{equation}
\vspace{-1.5mm}

\noindent where $\boldsymbol{\Psi}\in \mathbb{C}^{MN\times MN}$ contains channel delay and Doppler information. By defining $l_{\tau}\triangleq\lceil\tau M\Delta f\rceil$, the elements of $\boldsymbol{\Psi}$ are obtained by~\eqref{eq:Psi}, which is shown at the bottom of the page.
\endgroup
\stepcounter{equation}

\subsection{Multi-Target Sensing Model}
For simplicity, this study assumes each target is characterized by its line-of-sight (LoS) paths only. Suppose target $t$ is located at \begingroup\makeatletter\def\f@size{9.5} $\mathbf{p}_t=[x_t,y_t]^{\mathrm{T}}$ in the horizontal coordinate with a velocity of $\mathbf{v}_t$. Similarly, let $\mathbf{p}_p$ and $\mathbf{p}_r$ \endgroup denote the positions of the $p$th transmitting AP and the $r$th receiving AP, respectively. Then, the parameters of the reflected path from the $p$th transmitting AP via the $t$th target to the $r$th receiving AP, namely the angle of arrival (AoA), angle of departure (AoD), bi-static delay and Doppler, \begingroup\makeatletter\def\f@size{9.5}\check@mathfonts $\boldsymbol{\theta}_{p,r,t}^{(1)}\!\triangleq[\omega_{p,r,t}^{\mathrm{r}},\omega_{p,r,t}^{\mathrm{t}},\tau_{p,r,t},\nu_{p,r,t}]^\mathrm{T}\!,$ can be obtained by
\vspace{-1mm}
\begin{equation}
\begin{aligned}
\hspace {1.5pc}&\omega_{p,r,t}^{\mathrm{r}}=\pi\boldsymbol{\rho}_{tr}^{\mathrm{T}}\mathbf{u}_r^{\phantom{0}},\quad\tau_{p,r,t}=\left(d_{pt}+d_{tr}\right)/c,\\
&\omega_{p,r,t}^{\mathrm{t}}=\pi\boldsymbol{\rho}_{pt}^{\mathrm{T}}\mathbf{u}_p^{\phantom{0}},\quad\nu_{p,r,t}=\mathbf{v}_t^{\mathrm{T}}\left(\boldsymbol{\rho}_{pt}+\boldsymbol{\rho}_{tr}\right)/\lambda_c,
\end{aligned}
\end{equation}
\endgroup
\vspace{-1mm}

\noindent where $\lambda_{c}$ is the carrier wavelength; \begingroup\makeatletter\def\f@size{9.5}\check@mathfonts $d_{pt}=\|\mathbf{p}_p-\mathbf{p}_t\|$ and $d_{tr}=\|\mathbf{p}_t-\mathbf{p}_r\|$ \endgroup denote the distances from transmitting AP $p$ and receiving AP $r$ to the $t$th target location, respectively; $\mathbf{u}_p$ and $\mathbf{u}_r$ represent the unit direction vectors of the antenna elements at transmitting AP $p$ and receiving AP $r$, respectively~\cite{Gong2023Simultaneous}; and \begingroup\makeatletter\def\f@size{9.5}\check@mathfonts $\boldsymbol{\rho}_{pt}=(\mathbf{p}_p-\mathbf{p}_t)/d_{pt}$, $\boldsymbol{\rho}_{tr}=(\mathbf{p}_t-\mathbf{p}_r)/d_{tr}$ \endgroup are defined.


Assuming half-wavelength-spaced antennas on each AP, the antenna array response for an azimuth angle $\omega$ is given as
\vspace{-1mm}
\begingroup\makeatletter\def\f@size{9}\check@mathfonts
\begin{equation} \mathbf{a}(\omega)=\frac{1}{\sqrt{M_{\mathrm t}}}\left[ 1, e^{-j\omega}, \ldots, e^{-j(M_{\mathrm t}-1)\omega}\right]^\mathrm{T}\!\in\mathbb{C}^{M_{\mathrm t}\times1}.
\vspace{-1mm}
\end{equation}
\endgroup

\noindent Accordingly, the array response from the $p$th transmitting AP and the $r$th receiving AP to the $t$th target can be expressed by $\mathbf{h}_{pt}=\mathbf{a}(\omega_{p,r,t}^{\mathrm{t}})$ and $\mathbf{h}_{tr}=\mathbf{a}(\omega_{p,r,t}^{\mathrm{r}})$, respectively. Considering target location uncertainty and pointing the sensing beam for target $t$ to its approximate position with an AoD of $\hat{\omega}_{p,r,t}^{\mathrm{t}}$ yield
\begingroup\makeatletter\def\f@size{9}\check@mathfonts
\begin{equation} 
\vspace{-1mm}
\hat{\mathbf{h}}_{pt}=\mathbf{a}(\hat{\omega}_{p,r,t}^{\mathrm{t}}).\label{eq:sensing_BF}
\vspace{-1mm}
\end{equation}
\endgroup

This study adopts the assumption that the transmitted ISAC signal is available at the CPU, which indicates that the communication symbols also contribute to sensing via the reflected paths toward the targets~\cite{behdad2024multi}. Then, the received signal $\mathbf{y}_r\in\mathbb{C}^{M_\mathrm{t}MN\times1}$ at receiving AP $r$ is formulated as
\begingroup\makeatletter\def\f@size{9.5}\check@mathfonts
\begin{align}
\vspace{-1mm}
\hspace {-0.5pc}\mathbf{y}_r=\!\sum_{p=1}^{N_{\mathrm{tx}}}\sum_{t=1}^{T_{\mathrm{g}}}{\mathbf{H}}_{prt}^{\phantom{\dagger}}\bigg(\sum_{t^{\prime}=1}^{T_{\mathrm{g}}}\eta _{pt^{\prime}}^{\frac{1}{2}} \hat {\mathbf{H}}_{pt^{\prime}}^{\dagger}\mathbf{x}_{t^{\prime}}\!+\!\!\sum_{q=1}^{K_{\mathrm{u}}}\eta _{pq^{\phantom{\prime}}}^{\frac{1}{2}}\!\hat {\mathbf{H}}_{pq}^{\dagger}\mathbf{x}_{q}\!\bigg)\!+\!\mathbf{w}_r,\!
\label{eq:sensing_received}
\vspace{-1mm}
\end{align}

\noindent where $\mathbf{H}_{prt}\triangleq\beta_{p,r,t}\big(\mathbf{h}_{tr}^{\phantom{T}}\mathbf{h}_{pt}^\mathrm{T}\otimes \boldsymbol{\Psi}_{p,r,t}\big)$ denotes the sensing reflected channel. Here, $\beta_{p,r,t}\triangleq\alpha_{p,r,t}\xi_{p,r,t}^{1/2}$, where $\alpha_{p,r,t}\sim\mathcal{CN}(0,\sigma_{p,r,t}^{2})$ \endgroup is the radar cross-section (RCS) of the $t$th target, and $\xi_{p,r,t}=\frac{\lambda_{c}^{2}G_{\mathrm{t}}G_{\mathrm{r}}}{(4\pi)^{3}d_{pt}^{2}d_{tr}^{2}}$ with $G_{\mathrm{t}}$ and $G_{\mathrm{r}}$ denoting the antenna gains at the transmitting and receiving APs, respectively.
 

\section{CRLB Derivation}
The CRLB expresses a lower bound on the variance of unbiased estimators of deterministic parameters. To evaluate sensing performance, this section derives a closed-form CRLB for multi-target positioning error and proposes a low-complexity approximation.

\subsection{The Original CRLB Expression}
To decouple the channel gain from the other geometric channel parameters, this study represents the channel parameters from the $p$th transmitting AP via the $t$th target to the $r$th receiving AP as
\vspace{-1mm}
\begingroup\makeatletter\def\f@size{9.5}\check@mathfonts
\begin{align}
\boldsymbol{\theta}_{p,r,t}\triangleq\big[\big(\boldsymbol{\theta}_{p,r,t}^{(1)}\big)^\mathrm{T},\big(\boldsymbol{\theta}_{p,r,t}^{(2)}\big)^\mathrm{T}\big]^\mathrm{T}\in\mathbb{R}^{6\times1},
\label{eq:theta}
\end{align}

\noindent where $\boldsymbol{\theta}_{p,r,t}^{(2)}\triangleq[\beta_{p,r,t}^{(\mathrm{R})},\beta_{p,r,t}^{(\mathrm{I})}]^\mathrm{T}$ with $\beta_{p,r,t}^{(\mathrm{R})}=\Re\{\beta_{p,r,t}\}$ and $\beta_{p,r,t}^{(\mathrm{I})}=\Im\{\beta_{p,r,t}\} \vphantom{\underline{\beta_{p,r,t}^{(\mathrm{I})}}} $\endgroup. Let $\bar{\mathbf{y}}_{r}[k,l]$ denote the noiseless part of the received signal at DD grid $[k,l]$ in~\eqref{eq:sensing_received}, then the $(i,j)th$ element of the Fisher information matrix (FIM) concerning \begingroup\makeatletter\def\f@size{9.5}\check@mathfonts $\boldsymbol{\theta}_{p,r,t}$ can be computed by~\cite{Pucci2025Cooperative}
\vspace{-1mm}
\vspace{-1mm}
\begin{equation}
[\mathbf{F}_{\boldsymbol{\theta}_{p,r,t}}]_{i,j}=\frac{2}{\sigma_w^2}\Re\left\{\sum_{k=0}^{N-1}\sum_{l=0}^{M-1}\left(\frac{\partial\bar{\mathbf{y}}_{r}[k,l]}{\partial\boldsymbol{\theta}_{p,r,t}[i]}\right)^{\dagger}\!\left(\frac{\partial\bar{\mathbf{y}}_{r}[k,l]}{\partial\boldsymbol{\theta}_{p,r,t}[j]}\right)\right\},
\label{eq:FIM}
\vspace{-1mm}
\end{equation}
\endgroup


\noindent which has a form of~\cite{AbuShaban2018Error}\vspace{-1mm}
\begin{equation}
\hspace {-0.4pc}F_{x,x^{\prime}}\!=\!\Re\left\{(\mathrm{RX~factor})\!\times\!(\mathrm{TX~factor})\!\times\!(\text{signal factor})\right\}\!.\!
\label{eq:FIM_form}
\vspace{-1mm}
\end{equation}


\noindent For instance, it can be verified that
\vspace{-1mm}
\begingroup\makeatletter\def\f@size{9.5}\check@mathfonts
\begin{equation}
\hspace {-0.5pc}F_{\omega_{p,r,t}^{\mathrm{r}},\omega_{p,r,t}^{\mathrm{r}}}=\frac{2}{\sigma_w^2}\Re\{(\beta_{p}^{*}\beta_{p}^{\phantom{*}}\dot{\mathbf{h}}_{tr}^{\dagger}\dot{\mathbf{h}}_{tr}^{\phantom{\dagger}})(\mathbf{h}_{pt}^{\dagger}\mathbf{V}_{p}^{\phantom{\dagger}}\mathbf{h}_{pt}^{\phantom{\dagger}})R_{p,r,t}^{(0,0)}\}.
\label{eq:FIM_example}
\vspace{-1mm}
\end{equation}

\noindent where $\dot{\mathbf{h}}_{tr}=\mathbf{c}_{M_{\mathrm{t}}}\!\odot\mathbf{h}_{tr}$, in which $\mathbf{c}_{M_{\mathrm{t}}}\!=[0,1,\ldots,M_{\mathrm{t}}-1]^\mathrm{T}$; and $\mathbf{V}_{p}\triangleq\sum_{t=1}^{T_\mathrm{g}}\eta_{pt}\hat{\mathbf{h}}_{pt}^{\phantom{\dagger}}\hat{\mathbf{h}}_{pt}^{\dagger} + \sum_{q_{\vphantom{1}}=1}^{K_\mathrm{u}}\sum_{i=1}^{L_{pq}^{\vphantom 1}}\eta_{pq}\hat{\mathbf{h}}_{pq,i}^{\phantom{\dagger}}\hat{\mathbf{h}}_{pq,i}^{\dagger} \vphantom{\underline{\hat{\mathbf{h}}_{pq,i}^{\phantom{\dagger}}\hat{\mathbf{h}}_{pq,i}^{\dagger}}} $. The remaining entries of~\eqref{eq:FIM} exhibit the structure in~\eqref{eq:FIM_form}, which are provided in Appendix~\ref{app:FIM_Entries}.
\endgroup

In adherence to~\eqref{eq:theta}, the FIM can be partitioned as
\begingroup\makeatletter\def\f@size{9.5}\check@mathfonts
\begin{equation}
\vspace{-1mm}
\mathbf{F}_{\boldsymbol{\theta}_{p,r,t}} = \left[\begin{matrix}
\mathbf{F}_{\boldsymbol{\theta}_{p,r,t}^{(1)}} & \mathbf{F}_{\boldsymbol{\theta}_{p,r,t}^{(1)}, \boldsymbol{\theta}_{p,r,t}^{(2)}}  \\
\mathbf{F}_{\boldsymbol{\theta}_{p,r,t}^{(1)}, \boldsymbol{\theta}_{p,r,t}^{(2)}}^\mathrm{T} & \mathbf{F}_{\boldsymbol{\theta}_{p,r,t}^{(2)}}  \\
\end{matrix}\right]. 
\end{equation}
\vspace{-1mm}

\noindent Consequently, the equivalent FIM of $\boldsymbol{\theta}_{p,r,t}^{(1)}$, which includes only the parameters related to the target position, is given by
\vspace{-1mm}
\begin{equation}
\mathbf{F}_{\boldsymbol{\theta}_{p,r,t}^{(1)}}^{\mathrm{e}}\!=\mathbf{F}_{\boldsymbol{\theta}_{p,r,t}^{(1)}}\!-\mathbf{F}_{\boldsymbol{\theta}_{p,r,t}^{(1)}, \boldsymbol{\theta}_{p,r,t}^{(2)}} \mathbf{F}_{\boldsymbol{\theta}_{p,r,t}^{(2)}}^{-1} \mathbf{F}_{\boldsymbol{\theta}_{p,r,t}^{(1)}, \boldsymbol{\theta}_{p,r,t}^{(2)}}^{\mathrm{T}}\!\in\mathbb{R}^{4\times4}.
\label{eq:EFIM}
\vspace{-0.6mm}
\end{equation}

\noindent After collecting information from all reflected paths, the FIM for multi-target position estimation of target $t$ is given as
\vspace{-1mm}
\begin{equation}
\mathbf{F}_{\mathbf{p}_t}=\sum_{p=1}^{N_{\mathrm{tx}}}\sum_{r=1}^{N_{\mathrm{rx}}}\nabla_{\mathbf{p}_t}^{\mathrm{T}}\boldsymbol{\theta}_{p,r,t}^{(1)}\, \mathbf{F}_{\boldsymbol{\theta}_{p,r,t}^{(1)}}^{\mathrm{e}}\!\!\nabla_{\mathbf{p}_t}^{\phantom{\mathrm{T}}}\boldsymbol{\theta}_{p,r,t}^{(1)}\in\mathbb{R}^{2\times2},
\label{eq:positionFIM}
\vspace{-1mm}
\end{equation}
\endgroup
\vspace{-1mm}

\noindent where the Jacobian is expressed as
\vspace{-1mm}
\renewcommand{\arraystretch}{1.25}
\begin{align}
\nabla_{\mathbf{p}_t} \boldsymbol{\theta}_{p,r,t}^{(1)}=&\left[\!\begin{array}{c}
\pi\, \mathbf{u}_r^{\mathrm{T}}\left( \frac{\mathbf{I}-\boldsymbol{\rho}_{tr} \boldsymbol{\rho}_{tr}^{\mathrm{T}}}{\left\| \mathbf{p}_t-\mathbf{p}_r \right\|} \right) \\
\pi\, \mathbf{u}_p^{\mathrm{T}}\left( \frac{\mathbf{I}-\boldsymbol{\rho}_{pt} \boldsymbol{\rho}_{pt}^{\mathrm{T}}}{\left\| \mathbf{p}_t-\mathbf{p}_p \right\|} \right) \\
\frac{1}{c}\left(\boldsymbol{\rho}_{tr}+\boldsymbol{\rho}_{pt}\right)\!\!^{\mathrm{T}}\\
\frac{\mathbf{v}_t^{\mathrm{T}}}{\lambda_c}\left( \frac{\mathbf{I}-\boldsymbol{\rho}_{tr} \boldsymbol{\rho}_{tr}^{\mathrm{T}}}{\left\| \mathbf{p}_t-\mathbf{p}_r \right\|} + \frac{\mathbf{I}-\boldsymbol{\rho}_{pt} \boldsymbol{\rho}_{pt}^{\mathrm{T}}}{\left\| \mathbf{p}_t-\mathbf{p}_p \right\|} \right) \\
\end{array}\!\right]
\!\triangleq\!\left[\!\begin{array}{c}
\boldsymbol{J}_{1}^{\mathrm{T}} \\
\boldsymbol{J}_{2}^{\mathrm{T}} \\
\boldsymbol{J}_{3}^{\mathrm{T}} \\
\boldsymbol{J}_{4}^{\mathrm{T}} \\
\end{array}\!\right]
\vspace{-1mm}
\end{align}
\renewcommand{\arraystretch}{1}

Finally, the CRLB for the positioning error of target $t$ can then be obtained as $\text{CRLB}_{\mathbf{p}_t}=$ \begingroup\makeatletter\def\f@size{9.5}\check@mathfonts $\mathrm{Tr}\left( \mathbf{F}_{\mathbf{p}_t}^{-1} \right)$ \endgroup, and the PEB is defined as
\vspace{-1mm}
\begin{equation}
\text{PEB}_{\mathbf{p}_t}\triangleq\sqrt{\text{CRLB}_{\mathbf{p}_t}}.
\end{equation}
\vspace{-1mm}
\vspace{-1mm}
\vspace{-1mm}
\vspace{-1mm}
\vspace{-1mm}
\vspace{-1mm}

\subsection{Approximate Fisher Information}
Although the original FIM presents a closed-form expression for the calculation of sensing CRLB, it is critical to note that the computational complexity of this expression scales as $\mathcal{O}\big(M^{2}N^{2}N_{\mathrm{tx}}N_{\mathrm{rx}}T_{\mathrm{g}}\big)$. In practice, the high dimensionality of matrix $\boldsymbol{\Psi}_{p,r,t}$ exacerbates this complexity, significantly hindering CRLB analysis and optimization. Therefore, developing a low-complexity expression becomes essential for practical implementation.


\begin{proposition}\label{prop1}
By considering only the beam directed toward the corresponding target, an approximation of the FIM for multi-target position estimation of target $t$ in~\eqref{eq:positionFIM} can be obtained by
\begingroup\makeatletter\def\f@size{9.5}\check@mathfonts
$\mathbf{F}_{\mathbf{p}_t}=\sum_{p=1}^{N_{\mathrm{tx}}}\sum_{r=1}^{N_{\mathrm{rx}}}\eta_{pt}\hat{\mathbf{F}}_{\mathbf{p}_{p,r,t}}$, with \vspace{-1mm}
\begin{equation}
\hat{\mathbf{F}}_{\mathbf{p}_{p,r,t}}\!\approx\frac{2|\beta_{p}|^{2}}{\sigma_w^2}\left(d_{11}\boldsymbol{J}_{1}^{\phantom{\mathrm{T}}}\!\boldsymbol{J}_{1}^{\mathrm{T}}\!+d_{33}\boldsymbol{J}_{3}^{\phantom{\mathrm{T}}}\!\boldsymbol{J}_{3}^{\mathrm{T}}\!+d_{44}\boldsymbol{J}_{4}^{\phantom{\mathrm{T}}}\!\boldsymbol{J}_{4}^{\mathrm{T}}\right)\!,
\label{eq:approxPositionFIM}
\end{equation}
\endgroup
\end{proposition}
\vspace{-1mm}
\noindent where $d_{11}$, $d_{22}$ and $d_{33}$ are given in~\eqref{eq:EFIM_derived}.

\begin{IEEEproof}
  The proof is given in Appendix~\ref{app:prop1}.
\end{IEEEproof} 


\section{Proposed Power Allocation Algorithm}
In this study, communication SINR is used as a performance metric of the CF-ISAC system, given by~\cite{Fan2024Power}\vspace{-1mm}
\begingroup\makeatletter\def\f@size{9.5}\check@mathfonts
\begin{equation}
\hspace {-0.4pc}\text{SINR}^{(\mathrm c)}_{q}\!=\! \frac{\left( \sum_{p=1}^{N_{\mathrm{tx}}}{\eta _{pq}^{1/2}}b_{pq}\right) ^2}{\sum _{p=1}^{N_{\mathrm{tx}}}\Big(\sum _{q^{\prime}=0}^{K_{\mathrm{u}}}\eta_{pq^{\prime}}a_{pq,q^{\prime}}\!+\!\sum_{t=1}^{T_\mathrm{g}}{\eta _{pt}}a_{pq,t}\Big)\!+\!\sigma_w^2},\!\label{eq:SINR_c}
\end{equation}
\endgroup

{\noindent where {\small$ b_{pq}\!\triangleq\!{\sum_{i=1}^{L_{pq}}}\mathrm{Tr}\left( \mathbf{B}_{pq,i}\right)$}, {\small$a_{pq,t}\!\triangleq\!\sum_{i=1}^{L_{pq}}{\mathrm{Tr}\left(\mathbf{B}_{pq,i}\mathbf{B}_{pt}\right)}$} and {\small$a_{pq,q^{\prime}}\!\triangleq\!\sum_{i=1}^{L_{pq}^{\phantom{1}}}\sum_{j=1}^{L_{pq^{\prime}}^{\phantom{1}}}{\mathrm{Tr}\left(\mathbf{B}_{pq,i}\mathbf{B}_{pq^{\prime}\!,j}\right)}$} are defined. Here, {\small$\mathbf{B}_{pq,i} \triangleq \mathbb{E} \{\hat{\mathbf{h}}_{pq,i}^{\phantom{\dagger}}\hat{\mathbf{h}}_{pq,i}^{\dagger}\}$} denotes the covariance matrix of estimated channel vector {\small$\hat{\mathbf{h}}_{pq,i}$}. Please refer to~\cite{Fan2024Power} for a detailed explanation of\parfillskip=0pt\par}

\setlength{\textfloatsep}{0pt}

\noindent the channel estimation process. Meanwhile, by recalling~\eqref{eq:sensing_BF}, the sensing precoding matrix and its trace can be defined as {\small$\mathbf{B}_{pt}\triangleq\hat{\mathbf{h}}_{pt}^{\phantom{\dagger}}\hat{\mathbf{h}}_{pt}^{\dagger}$} and {\small$b_{pt}\!\triangleq\!\mathrm{Tr}\left( \mathbf{B}_{pt}\right)$}.

This section maximizes the communication SINR, the max-min fairness optimization problem can be expressed as follows:
\vspace{-12.5pt}
\begin{subequations}\label{eq:35}
\begin{alignat}{2}
& \underset{\boldsymbol{\eta}\geq\mathbf{0}}{\text{maximize}} &\quad& \underset{q\in\{1,...,K_{\mathrm u}\}}{\text{min}}\Big\{\text{SINR}^{(\mathrm c)}_{q}\Big\} \label{eq:35a}
\\ 
& \text{subject to} && \text{CRLB}_{\mathbf{p}_t} \leq \gamma_{\mathrm s},\quad t=1,\ldots,T_\mathrm{g}, \label{eq:35b}
\\
& && P_{p}\leq P_{\mathrm{d}},\quad\quad\,\quad p=1,\ldots,N_{\mathrm{tx}}, \label{eq:35c}
\end{alignat}
\end{subequations}

\noindent where $\boldsymbol{\eta}\in\mathbb{R}^{N_{\mathrm{tx}}(K_{\mathrm{u}}+T_{\mathrm{g}})\times1}$ is the concatenated power allocation coefficient vector; $\gamma_{\mathrm s}$ denotes the maximum sensing CRLB threshold, and~\eqref{eq:35c} is the power constraint defined by~\eqref{eq:power_constraint}.


\begin{figure*}[t]
  \centering
  \begin{minipage}[t]{0.325\linewidth}
    \centering
    \includegraphics[width=2.2in]{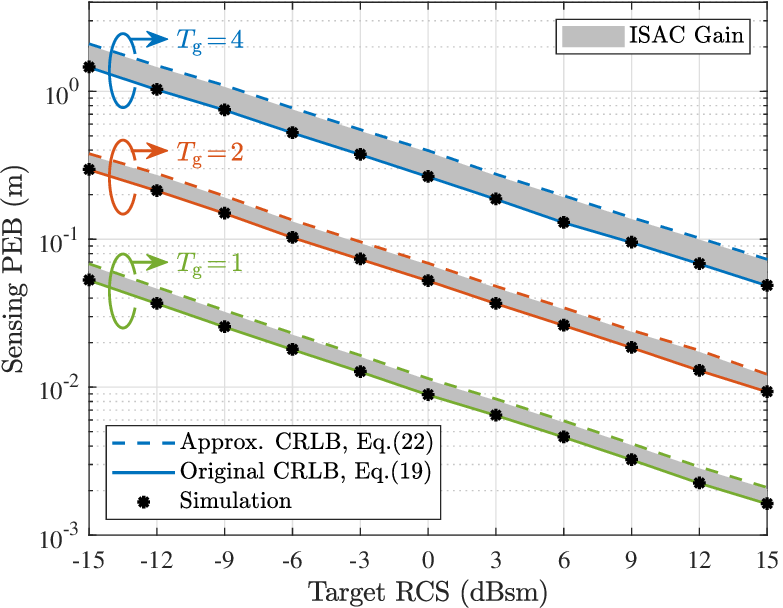}
\vspace{-0.3cm}
    \caption{The sensing PEB versus different target number and RCS variance.}
      \label{fig:fig_2}
\vspace{-0.3cm}
  \end{minipage}%
    \hspace{0.3cm}
  \begin{minipage}[t]{0.30\linewidth}
    \centering
    \includegraphics[width=2.12in]{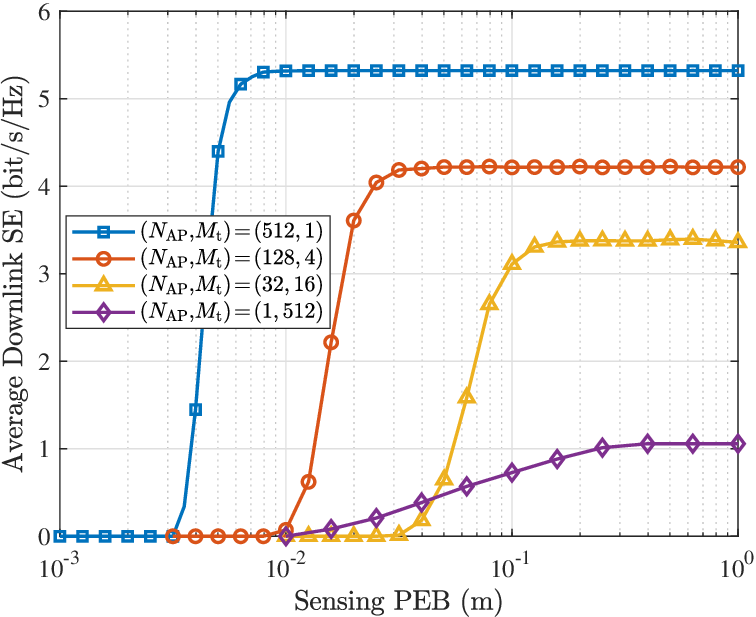}
\vspace{-0.3cm}
    \caption{Tradeoff between the SE and the sensing PEB constraint in both cellular and CF systems.}
      \label{fig:fig_3}
\vspace{-0.3cm}
  \end{minipage}
    \hspace{0.3cm}
  \begin{minipage}[t]{0.30\linewidth}
    \centering
    \includegraphics[width=2.2in]{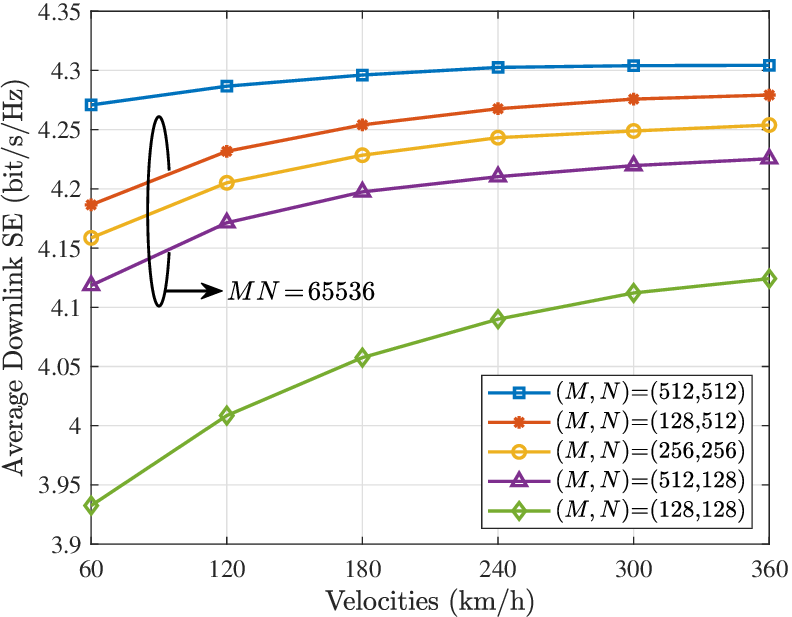}
\vspace{-0.7cm}
    \caption{The average per-user communication SE versus different user and target velocities.}
      \label{fig:fig_4}
  \end{minipage}
\vspace{-0.3cm}
\end{figure*}

\begin{lemma}\label{lem1}
The power constraint in~\eqref{eq:35c} at the $p$th transmitting AP can be obtained by
\vspace{-1mm}
\begingroup\makeatletter\def\f@size{9}\check@mathfonts
\begin{equation}\label{eq:power_constraint_new}
P_{p}=\sum _{q=1}^{K_{\mathrm{u}}}\eta _{pq}b_{pq}+\sum _{t=1}^{T_{\mathrm{g}}}\eta _{pt}b_{pt}\leq P_{\mathrm{d}}.
\end{equation}
\endgroup
\end{lemma}
\vspace{-1mm}
\begin{IEEEproof}
  The proof is similar to that in~\cite[Appendix C]{Fan2024Power}, and is therefore omitted due to space limitation.
\end{IEEEproof}

\begin{small}
\begin{algorithm}[!t]
\caption{Power Allocation Scheme for Multi-Target ISAC}
\label{alg1}
\begin{algorithmic}[1]
\STATE $\textbf{Initialization:}$ Set an arbitrary initial positive $\boldsymbol{\eta}^{(0)}$, the tolerance $\epsilon>0$ and the maximum iteration number $I$. Set the iteration counter to $t=0$ and $z^{(0)}=0$.
\REPEAT
\STATE $t\leftarrow t+1$.
    \STATE Update $y_q^{(t)}$ according to \eqref{eq:opt_yq};
    \STATE Update $\boldsymbol{\eta}^{(t)}$ by solving the convex optimization problem \eqref{eq:opt_main} for fixed $y_q$;
\UNTIL $|z^{(t)}-z^{(t-1)}|\leq\epsilon$ \OR $t=I$.
\STATE $\textbf{Output:}$ The transmit power coefficients $\boldsymbol{\eta}^{(t)}$.
\end{algorithmic}
\end{algorithm}
\end{small}


Further, since the trace of the inverse, {\small$\mathrm{Tr}\left( \mathbf{X}^{-1} \right)$}, is convex, it can be verified that the objective function in~\eqref{eq:SINR_c} exhibits a fractional programming structure, whereas the constraints \eqref{eq:35b}, \eqref{eq:power_constraint_new} are convex. Therefore, by applying the quadratic transform~\cite{shen2018fractional}, the original optimization problem~\eqref{eq:35} can be reformulated as a convex optimization problem, which can be expressed as
\begingroup\makeatletter\def\f@size{9}\check@mathfonts
\begin{subequations}
\vspace{-1mm}
\begin{alignat}{2}
& \underset{\boldsymbol{\eta}\geq\mathbf{0},~z}{\text{maximize}} &\ \ \!& z 
\\ 
& \text{subject to} && \mathrm{Tr}\left( \mathbf{F}_{\mathbf{p}_t}^{-1} \right) \leq \gamma_{\mathrm s},\quad t=1,\ldots,T_\mathrm{g},
\\
& && \sum _{q=1}^{K_{\mathrm{u}}}\eta _{pq}b_{pq}+\!\sum _{t=1}^{T_{\mathrm{g}}}\eta _{pt}b_{pt}\leq P_{\mathrm{d}},\quad\!\! p=1,\ldots,N_\mathrm{tx},\!
\\
& && -y_q^2\bigg(\!\sum_{p=1}^{N_{\mathrm{tx}}}\sum_{q^{\prime}=1}^{K_{\mathrm{u}}}{\eta _{pq^{\prime}}}a_{pq,q^{\prime}}\!+\!\sum_{p=1}^{N_{\mathrm{tx}}}\sum_{t=1}^{T_\mathrm{g}}{\eta _{pt}}a_{pq,t}+\!\sigma_w^2\!\bigg)\!\!\nonumber \\
& && \quad\quad\ \,\!+2y_q\sum_{p=1}^{N_{\mathrm{tx}}}\!\eta _{pq}^{1/2}b_{pq}\geq z,\quad\, q=1,\ldots,K_{\mathrm u},\!
\end{alignat}
\label{eq:opt_main}
\end{subequations}
\endgroup
\vspace{-10pt}

\noindent where the auxiliary variable $y_q$ for fixed $\boldsymbol{\eta}$ is defined as
\begin{equation} 
\vspace{-1mm}
y_q=\frac{\sum _{p=1}^{N_{\mathrm{tx}}}\eta_{pq}^{1/2}b_{pq}}{\sum _{p=1}^{N_{\mathrm{tx}}}\Big(\sum _{q^{\prime}=0}^{K_{\mathrm{u}}}\eta_{pq^{\prime}}a_{pq,q^{\prime}}\!+\!\sum_{t=1}^{T_\mathrm{g}}{\eta _{pt}}a_{pq,t}\Big)\!+\!\sigma_w^2}.
\label{eq:opt_yq}
\end{equation}
\noindent The optimization problem can be solved through an iterative approach, as outlined in Algorithm~\ref{alg1}. The computational complexity of this method is {\small$\mathcal{O}\left(I\left(N_{\mathrm{tx}}(K_{\mathrm{u}}+T_\mathrm{g})\right)^{3.5}\right)$}, where {\small$I$} denotes the maximum number of iterations.

\section{Numerical Results}
This section presents a comprehensive numerical analysis to evaluate the performance of applying the OTFS signals to the CF-ISAC systems. The key simulation parameters are listed in Table~\ref{tab3} unless stated otherwise. The path loss for the communication and sensing channels are modeled by the 3GPP Urban Microcell model and radar equation, respectively. A total number of $N_{\mathrm{AP}}$ APs, $K_{\mathrm u}$ users, and $T_{\mathrm{g}}$ targets are randomly distributed in a 300\,m\,$\times$\,300\,m area. For each target, the $N_{\mathrm{rx}}=2$ closest APs are assigned as sensing receivers, with the remaining $N_\mathrm{tx}$ APs serving as ISAC transmitters~\cite{behdad2024multi}.

\begin{table}[!t]
\vspace{-5pt}
\centering
\caption{Simulation Parameters}
\vspace{-3pt}
\label{tab3}
\rowcolors{2}{lightgray!30}{white}
\renewcommand\arraystretch{1.45}{
    \centering
    \begin{tabular}{ccc}
    \Xhline{0.8pt}\rowcolor{gray!35}
        \textbf{Parameters} & \textbf{Symbol} & \textbf{Value} \\ \Xhline{0.5pt}
        Carrier frequency & $f_{c}$ & 38\,GHz \\ 
        Bandwidth & $B$ & 64\,MHz \\ 
        Number of subcarriers & $M$ & 128 \\  
        Number of symbols & $N$ & 128 \\  
        Scenario size & - & 300m\,$\times$\,300m \\   
        Number of APs & $N_{\mathrm{AP}}$ & 32 \\  
        Number of antennas at each AP & $M_{\mathrm t}$ & 16 \\  
        Number of users & $K_{\mathrm u}$ & 10 \\ 
        Number of targets & $T_{\mathrm{g}}$ & 2 \\ 
        Maximum speed $($UE$/$Target$)$ & $v_{\max}$ & 300\,km/h \\  
        CP sample length & $N_{\mathrm{cp}}$ & $\lceil\tau_{\max}M\Delta f\rceil$ \\ 
        Sensing PEB threshold & $\gamma_{\mathrm s}$ & 0.1\,m \\ 
        Maximum transmit power & $P_{\mathrm{d}}$ & 1\,W \\ 
        Noise variance & $\sigma_w^2$ & -89\,dBm \\  
        RCS variance & $\sigma_{rcs}^2$ & 0\,dBsm \\ \Xhline{0.8pt}
    \end{tabular}}
    \vspace{5pt}
\end{table}

Fig.~\ref{fig:fig_2} presents the analytical sensing PEB calculated by~\eqref{eq:positionFIM} and the approximation in~\eqref{eq:approxPositionFIM}, along with the corresponding simulated results. Due to the high computational complexity of the original expression, transmit power optimization is not considered. Instead, the APs transmit with equal power, and the power control coefficients at the $p$th transmitting AP are set to \begingroup\makeatletter\def\f@size{9}\check@mathfonts $\eta_{pq}=\eta_{pt}=P_{\mathrm{d}}/\big(\sum _{q=1}^{K_{\mathrm{u}}}b_{pq}+\!\sum _{t=1}^{T_{\mathrm{g}}}b_{pt}\big)$. \endgroup Since the effects of the communication beams and the sensing beams directed at other targets are not considered, the proposed approximation serves as an upper bound of the original PEB, with their gap representing the ISAC coordination gain. Further, it is observed that the positioning accuracy degrades as the number of simultaneously sensed targets increases, in which case ISAC becomes more beneficial with an enhanced coordination gain.

The tradeoff between the communication SE and the sensing PEB constraint under different antenna configurations is illustrated in Fig.~\ref{fig:fig_3}. For a fair comparison, the total number of antennas $N_{\mathrm{AP}}M_{\mathrm t}=512$ remains fixed. The results indicate that relaxing the sensing PEB constraints leads to an increase in average SE. Moreover, compared to conventional cellular ISAC with $M_{\mathrm t}=\text{512}$, the CF-ISAC system significantly enhances SE performance, as denser AP deployments shorten the distances to both users and targets, thereby mitigating signal fading for both communication and sensing.

Fig.~\ref{fig:fig_4} investigates the system performance under varying user and target velocities for different values of $M$ and $N$. It can be noted that the SE performance gradually improves as velocity increases. The underlying reason is that more distinct paths can be resolved in the Doppler domain as velocity increases, leading to better system performance. The results also show that the SE performance degrades as $M$ and $N$ decrease, with the reduction in $N$ having a more significant impact. This is attributed to the diversity loss caused by coarser DD grid resolution, and the $N$-associated Doppler resolution plays a more critical role in positioning under the simulation settings.

\section{Conclusion}
This paper studies the multi-target position estimation in the CF-ISAC systems employing the OTFS signal, concurrently analyzing and optimizing the system performance. A closed-form CRLB expression and its low-complexity approximation for target position estimation is derived, which is universally applicable to multi-static sensing systems. Moreover, a power allocation algorithm is proposed to maximize the lowest user SE while ensuring the sensing PEB constraint to enhance the joint performance. The numerical results validate the proposed PEB expression and approximation, which clearly shows the coordination gain of ISAC signals and paves the way for system optimization. In addition, the tradeoff between the communication and sensing metrics is elucidated with different antenna arrangements, providing deeper insights into building future architectures of ISAC systems.

{\appendices
\section{Entries of the FIM in \texorpdfstring{\eqref{eq:FIM}}{\ref{eq:FIM}}}
\label{app:FIM_Entries}
\vspace{-6mm}
\begingroup\makeatletter\def\f@size{9.5}\check@mathfonts
\begin{align*}
F_{\omega_{p,r,t}^{\mathrm{t}},\omega_{p,r,t}^{\mathrm{t}}}&=\frac{2}{\sigma_w^2}\Re\{(\beta_{p}^{*}\beta_{p}^{\phantom{*}}\mathbf{h}_{tr}^{\dagger}\mathbf{h}_{tr}^{\phantom{\dagger}})(\dot{\mathbf{h}}_{pt}^{\dagger}\mathbf{V}_{p}^{\phantom{\dagger}}\dot{\mathbf{h}}_{pt}^{\phantom{\dagger}})R_{p,r,t}^{(0,0)}\}\\
F_{\tau_{p,r,t},\tau_{p,r,t}}&=\frac{2}{\sigma_w^2}\Re\{(\beta_{p}^{*}\beta_{p}^{\phantom{*}}\mathbf{h}_{tr}^{\dagger}\mathbf{h}_{tr}^{\phantom{\dagger}})(\mathbf{h}_{pt}^{\dagger}\mathbf{V}_{p}^{\phantom{\dagger}}\mathbf{h}_{pt}^{\phantom{\dagger}})R_{p,r,t}^{(2,0)}\}\\
F_{\nu_{p,r,t},\nu_{p,r,t}}&=\frac{2}{\sigma_w^2}\Re\{(\beta_{p}^{*}\beta_{p}^{\phantom{*}}\mathbf{h}_{tr}^{\dagger}\mathbf{h}_{tr}^{\phantom{\dagger}})(\mathbf{h}_{pt}^{\dagger}\mathbf{V}_{p}^{\phantom{\dagger}}\mathbf{h}_{pt}^{\phantom{\dagger}})R_{p,r,t}^{(0,2)}\}\\
F_{\beta_{p,r,t}^{(\mathrm{R})},\beta_{p,r,t}^{(\mathrm{R})}}&=\frac{2}{\sigma_w^2}\Re\{(\mathbf{h}_{tr}^{\dagger}\mathbf{h}_{tr}^{\phantom{\dagger}})(\mathbf{h}_{pt}^{\dagger}\mathbf{V}_{p}^{\phantom{\dagger}}\mathbf{h}_{pt}^{\phantom{\dagger}})R_{p,r,t}^{(0,0)}\}\\
F_{\beta_{p,r,t}^{(\mathrm{I})},\beta_{p,r,t}^{(\mathrm{I})}}&=\frac{2}{\sigma_w^2}\Re\{(\mathbf{h}_{tr}^{\dagger}\mathbf{h}_{tr}^{\phantom{\dagger}})(\mathbf{h}_{pt}^{\dagger}\mathbf{V}_{p}^{\phantom{\dagger}}\mathbf{h}_{pt}^{\phantom{\dagger}})R_{p,r,t}^{(0,0)}\}\\
F_{\omega_{p,r,t}^{\mathrm{r}},\omega_{p,r,t}^{\mathrm{t}}}&=-\frac{2}{\sigma_w^2}\Re\{(\beta_{p}^{*}\beta_{p}^{\phantom{*}}\dot{\mathbf{h}}_{tr}^{\dagger}\mathbf{h}_{tr}^{\phantom{\dagger}})(\dot{\mathbf{h}}_{pt}^{\dagger}\mathbf{V}_{p}^{\phantom{\dagger}}\mathbf{h}_{pt}^{\phantom{\dagger}})R_{p,r,t}^{(0,0)}\}\\
F_{\omega_{p,r,t}^{\mathrm{r}},\tau_{p,r,t}}&=\frac{2}{\sigma_w^2}\Re\{j(\beta_{p}^{*}\beta_{p}^{\phantom{*}}\dot{\mathbf{h}}_{tr}^{\dagger}\mathbf{h}_{tr}^{\phantom{\dagger}})(\mathbf{h}_{pt}^{\dagger}\mathbf{V}_{p}^{\phantom{\dagger}}\mathbf{h}_{pt}^{\phantom{\dagger}})R_{p,r,t}^{(1,0)}\}\\
F_{\omega_{p,r,t}^{\mathrm{r}},\nu_{p,r,t}}&=\frac{2}{\sigma_w^2}\Re\{j(\beta_{p}^{*}\beta_{p}^{\phantom{*}}\dot{\mathbf{h}}_{tr}^{\dagger}\mathbf{h}_{tr}^{\phantom{\dagger}})(\mathbf{h}_{pt}^{\dagger}\mathbf{V}_{p}^{\phantom{\dagger}}\mathbf{h}_{pt}^{\phantom{\dagger}})R_{p,r,t}^{(0,1)}\}\\
F_{\omega_{p,r,t}^{\mathrm{r}},\beta_{p,r,t}^{(\mathrm{R})}}&=\frac{2}{\sigma_w^2}\Re\{j(\beta_{p}^{*}\dot{\mathbf{h}}_{tr}^{\dagger}\mathbf{h}_{tr}^{\phantom{\dagger}})(\mathbf{h}_{pt}^{\dagger}\mathbf{V}_{p}^{\phantom{\dagger}}\mathbf{h}_{pt}^{\phantom{\dagger}})R_{p,r,t}^{(0,0)}\}\\
F_{\omega_{p,r,t}^{\mathrm{r}},\beta_{p,r,t}^{(\mathrm{I})}}&=-\frac{2}{\sigma_w^2}\Re\{(\beta_{p}^{*}\dot{\mathbf{h}}_{tr}^{\dagger}\mathbf{h}_{tr}^{\phantom{\dagger}})(\mathbf{h}_{pt}^{\dagger}\mathbf{V}_{p}^{\phantom{\dagger}}\mathbf{h}_{pt}^{\phantom{\dagger}})R_{p,r,t}^{(0,0)}\}\\
F_{\omega_{p,r,t}^{\mathrm{t}},\tau_{p,r,t}}&=-\frac{2}{\sigma_w^2}\Re\{j(\beta_{p}^{*}\beta_{p}^{\phantom{*}}\mathbf{h}_{tr}^{\dagger}\mathbf{h}_{tr}^{\phantom{\dagger}})(\mathbf{h}_{pt}^{\dagger}\mathbf{V}_{p}^{\phantom{\dagger}}\dot{\mathbf{h}}_{pt}^{\phantom{\dagger}})R_{p,r,t}^{(1,0)}\}\\
F_{\omega_{p,r,t}^{\mathrm{t}},\nu_{p,r,t}}&=-\frac{2}{\sigma_w^2}\Re\{j(\beta_{p}^{*}\beta_{p}^{\phantom{*}}\mathbf{h}_{tr}^{\dagger}\mathbf{h}_{tr}^{\phantom{\dagger}})(\mathbf{h}_{pt}^{\dagger}\mathbf{V}_{p}^{\phantom{\dagger}}\dot{\mathbf{h}}_{pt}^{\phantom{\dagger}})R_{p,r,t}^{(0,1)}\}\\
F_{\omega_{p,r,t}^{\mathrm{t}},\beta_{p,r,t}^{(\mathrm{R})}}&=-\frac{2}{\sigma_w^2}\Re\{j(\beta_{p}^{*}\mathbf{h}_{tr}^{\dagger}\mathbf{h}_{tr}^{\phantom{\dagger}})(\mathbf{h}_{pt}^{\dagger}\mathbf{V}_{p}^{\phantom{\dagger}}\dot{\mathbf{h}}_{pt}^{\phantom{\dagger}})R_{p,r,t}^{(0,0)}\}\\
F_{\omega_{p,r,t}^{\mathrm{t}},\beta_{p,r,t}^{(\mathrm{I})}}&=\frac{2}{\sigma_w^2}\Re\{(\beta_{p}^{*}\mathbf{h}_{tr}^{\dagger}\mathbf{h}_{tr}^{\phantom{\dagger}})(\mathbf{h}_{pt}^{\dagger}\mathbf{V}_{p}^{\phantom{\dagger}}\dot{\mathbf{h}}_{pt}^{\phantom{\dagger}})R_{p,r,t}^{(0,0)}\}\\
F_{\tau_{p,r,t},\nu_{p,r,t}}&=\frac{2}{\sigma_w^2}\Re\{(\beta_{p}^{*}\beta_{p}^{\phantom{*}}\mathbf{h}_{tr}^{\dagger}\mathbf{h}_{tr}^{\phantom{\dagger}})(\mathbf{h}_{pt}^{\dagger}\mathbf{V}_{p}^{\phantom{\dagger}}\mathbf{h}_{pt}^{\phantom{\dagger}})R_{p,r,t}^{(1,1)}\}\\
F_{\tau_{p,r,t},\beta_{p,r,t}^{(\mathrm{R})}}&=\frac{2}{\sigma_w^2}\Re\{(\beta_{p}^{*}\mathbf{h}_{tr}^{\dagger}\mathbf{h}_{tr}^{\phantom{\dagger}})(\mathbf{h}_{pt}^{\dagger}\mathbf{V}_{p}^{\phantom{\dagger}}\mathbf{h}_{pt}^{\phantom{\dagger}})R_{p,r,t}^{*(1,0)}\}\\
F_{\tau_{p,r,t},\beta_{p,r,t}^{(\mathrm{I})}}&=\frac{2}{\sigma_w^2}\Re\{j(\beta_{p}^{*}\mathbf{h}_{tr}^{\dagger}\mathbf{h}_{tr}^{\phantom{\dagger}})(\mathbf{h}_{pt}^{\dagger}\mathbf{V}_{p}^{\phantom{\dagger}}\mathbf{h}_{pt}^{\phantom{\dagger}})R_{p,r,t}^{*(1,0)}\}\\
F_{\nu_{p,r,t},\beta_{p,r,t}^{(\mathrm{R})}}&=\frac{2}{\sigma_w^2}\Re\{(\beta_{p}^{*}\mathbf{h}_{tr}^{\dagger}\mathbf{h}_{tr}^{\phantom{\dagger}})(\mathbf{h}_{pt}^{\dagger}\mathbf{V}_{p}^{\phantom{\dagger}}\mathbf{h}_{pt}^{\phantom{\dagger}})R_{p,r,t}^{*(0,1)}\}\\
F_{\nu_{p,r,t},\beta_{p,r,t}^{(\mathrm{I})}}&=\frac{2}{\sigma_w^2}\Re\{j(\beta_{p}^{*}\mathbf{h}_{tr}^{\dagger}\mathbf{h}_{tr}^{\phantom{\dagger}})(\mathbf{h}_{pt}^{\dagger}\mathbf{V}_{p}^{\phantom{\dagger}}\mathbf{h}_{pt}^{\phantom{\dagger}})R_{p,r,t}^{*(0,1)}\}\\
F_{\beta_{p,r,t}^{(\mathrm{R})},\beta_{p,r,t}^{(\mathrm{I})}}&=\frac{2}{\sigma_w^2}\Re\{j(\mathbf{h}_{tr}^{\dagger}\mathbf{h}_{tr}^{\phantom{\dagger}})(\mathbf{h}_{pt}^{\dagger}\mathbf{V}_{p}^{\phantom{\dagger}}\mathbf{h}_{pt}^{\phantom{\dagger}})R_{p,r,t}^{(0,0)}\}
\end{align*}
\endgroup

\section{Proof of Proposition \ref{prop1}}
\label{app:prop1}

Before deriving the approximation FIM {\small$\hat{\mathbf{F}}_{\mathbf{p}_{p,r,t}}$}, we define the short notations used in~\eqref{eq:FIM_example} as follows
\begingroup\makeatletter\def\f@size{9}\check@mathfonts
\begin{subequations}
\begin{align}
R_{p,r,t}^{(0,0)}\!=&\!\sum_{k=0}^{N-1}\sum_{l=0}^{M-1}\sum_{k^{\prime}=0}^{N-1} \sum_{l^{\prime}=0}^{M-1}  \left(\Psi_{k,k^{\prime},l,l^{\prime}}^{p,r,t}\right)^{*}\Psi_{k,k^{\prime},l,l^{\prime}}^{p,r,t}\!\label{eq:R00}\\
R_{p,r,t}^{(1,0)}\!=&\!\sum_{k=0}^{N-1}\sum_{l=0}^{M-1}\sum_{k^{\prime}=0}^{N-1} \sum_{l^{\prime}=0}^{M-1}  \left(\Psi_{k,k^{\prime},l,l^{\prime}}^{p,r,t}\right)^{*}\frac{\partial \Psi_{k,k^{\prime},l,l^{\prime}}^{p,r,t}}{\partial \tau_{p,r,t}}\!\label{eq:R10}\\
R_{p,r,t}^{(0,1)}\!=&\!\sum_{k=0}^{N-1}\sum_{l=0}^{M-1}\sum_{k^{\prime}=0}^{N-1} \sum_{l^{\prime}=0}^{M-1}  \left(\Psi_{k,k^{\prime},l,l^{\prime}}^{p,r,t}\right)^{*}\frac{\partial \Psi_{k,k^{\prime},l,l^{\prime}}^{p,r,t}}{\partial \nu_{p,r,t}}\!\\
R_{p,r,t}^{(1,1)}\!=&\!\sum_{k=0}^{N-1}\sum_{l=0}^{M-1}\sum_{k^{\prime}=0}^{N-1} \sum_{l^{\prime}=0}^{M-1}  \left(\frac{\partial \Psi_{k,k^{\prime},l,l^{\prime}}^{p,r,t}}{\partial \tau_{p,r,t}}\right)^{*}\frac{\partial \Psi_{k,k^{\prime},l,l^{\prime}}^{p,r,t}}{\partial \nu_{p,r,t}}\!
\end{align}

\begin{align}
R_{p,r,t}^{(2,0)}\!=&\!\sum_{k=0}^{N-1}\sum_{l=0}^{M-1}\sum_{k^{\prime}=0}^{N-1} \sum_{l^{\prime}=0}^{M-1}  \left(\frac{\partial \Psi_{k,k^{\prime},l,l^{\prime}}^{p,r,t}}{\partial \tau_{p,r,t}}\right)^{*}\frac{\partial \Psi_{k,k^{\prime},l,l^{\prime}}^{p,r,t}}{\partial \tau_{p,r,t}}\!\\
R_{p,r,t}^{(0,2)}\!=&\!\sum_{k=0}^{N-1}\sum_{l=0}^{M-1}\sum_{k^{\prime}=0}^{N-1} \sum_{l^{\prime}=0}^{M-1}  \left(\frac{\partial \Psi_{k,k^{\prime},l,l^{\prime}}^{p,r,t}}{\partial \nu_{p,r,t}}\right)^{*}\frac{\partial \Psi_{k,k^{\prime},l,l^{\prime}}^{p,r,t}}{\partial \nu_{p,r,t}}\!\label{eq:R02}
\end{align}
\end{subequations}
\endgroup
\vspace{-1mm}
\vspace{-1mm}

\noindent where the partial derivatives of $\boldsymbol{\Psi}$ with respect to the channel parameters $\tau$ and $\nu$, ignoring indices $p$, $r$ and $t$, are given by
\vspace{-1mm}
\begingroup\makeatletter\def\f@size{9.5}\check@mathfonts
\begin{equation}
\begin{aligned}
\frac{\partial \Psi_{k,k^{\prime},l,l^{\prime}}}{\partial \tau}=&\frac{j 2 \pi \Delta f}{N M} \mathbf{1}_N^\mathrm{T} \boldsymbol{\alpha}_{k, k^{\prime}}(\nu) \mathbf{c}_M^\mathrm{T} \boldsymbol{\beta}_{k^{\prime},l,l^{\prime}}(\nu, \tau),\\
\frac{\partial \Psi_{k,k^{\prime},l,l^{\prime}}}{\partial \nu}=&\frac{j 2 \pi}{N M}\big[T \mathbf{c}_N^\mathrm{T} \boldsymbol{\alpha}_{k,k^{\prime}}(\nu) \mathbf{1}_M^\mathrm{T} \boldsymbol{\beta}_{k^{\prime},l,l^{\prime}}(\nu, \tau) \\& 
+ g(l) \mathbf{1}_N^\mathrm{T} \boldsymbol{\alpha}_{k,k^{\prime}}(\nu) \mathbf{1}_M^\mathrm{T} \boldsymbol{\beta}_{k^{\prime},l,l^{\prime}}(\nu, \tau)\big],
\end{aligned}
\end{equation}
\endgroup
\vspace{-1mm}
\vspace{-1mm}

\begingroup\makeatletter\def\f@size{9}\check@mathfonts
\noindent where $g(l)\!=\!\frac{l}{M\Delta f}$ for $l\!\in\!\mathcal{L}_{\mathrm{ICI}}(\tau)$ and $g(l)\!=\!\frac{l}{M\Delta f}-T$ for $l\!\in\!\mathcal{L}_{\mathrm{ISI}}(\tau)$; meanwhile, the vectors $\boldsymbol{\alpha}_{k,k^{\prime}}\!$ and $\boldsymbol{\beta}_{k^{\prime},l,l^{\prime}}\!$ are defined as $\boldsymbol{\alpha}_{k,k^{\prime}}(\nu)\!=\![\alpha_{0,k,k^{\prime}}(\nu),\ldots,\alpha_{N-1,k,k^{\prime}}(\nu)]^\mathrm{T}\!\in\!\mathbb{C}^{N\times1}$ and $\boldsymbol{\beta}_{k^{\prime},l,l^{\prime}}(\nu,\tau)\!=\![\beta_{0,k^{\prime},l,l^{\prime}}(\nu,\tau),\ldots,\beta_{M-1,k^{\prime},l,l^{\prime}}(\nu,\tau)]^\mathrm{T}\!\in\!\mathbb{C}^{M\times1}$, respectively.
\endgroup

\begingroup\makeatletter\def\f@size{9.5}\check@mathfonts
Next, we proceed with the derivation of $R_{p,r,t}^{(0,0)}$. By substituting~\eqref{eq:Psi} into~\eqref{eq:R00}, the $R_{p,r,t}^{(0,0)}$ can be derived as
\vspace{-1mm}
\begin{subequations}\label{eq:R_derived}
\begin{align}
\hspace {-0.5pc}R_{p,r,t}^{(0,0)}=&\frac{1}{M^2}\!\sum_{l=0}^{M-1}\sum_{l^{\prime}=0}^{M-1}\sum_{m^{\prime}=0}^{M-1}\sum_{m^{\prime\prime}=0}^{M-1}\!e^{-j2\pi\frac{l^{\prime}-l+\tau M\Delta f}{M}(m^{\prime}\!-m^{\prime\prime})}\!\!\nonumber\\
&\times\frac{1}{N^2}\!\sum_{k=0}^{N-1}\sum_{k^{\prime}=0}^{N-1}\sum_{n^{\prime}=0}^{N-1}\sum_{n^{\prime\prime}=0}^{N-1}\!e^{-j2\pi\frac{k^{\prime}-k+\nu NT}{N}(n^{\prime}\!-n^{\prime\prime})}\!\!\nonumber\\
\stackrel {(a)}{=}&\frac{1}{M^2N^2}\times M^3N^3=MN,
\end{align}
\vspace{-1mm}
\vspace{-1mm}
\vspace{-1mm}
\vspace{-1mm}
\vspace{-1mm}

{\noindent where in (a), we note that the sum is nonzero only when $m^{\prime}\!-m^{\prime\prime}=0$ and $n^{\prime}\!-n^{\prime\prime}=0$. The remaining terms~\eqref{eq:R10}-\eqref{eq:R02} can\parfillskip=0pt\par}

\noindent be similarly obtained as follows
\vspace{-1mm}
\begin{align}
\hspace {-0.5pc}R_{p,r,t}^{(1,0)}\!=&j\pi\Delta f\left(M\!-\!1\right)MN\\
\hspace {-0.5pc}R_{p,r,t}^{(0,1)}\!=&j\pi\left[T\left(N\!-\!1\right)MN+2N\!\sum\nolimits_{l=0}^{M-1}\!\!g(l)\right]\\		
\hspace {-0.5pc}R_{p,r,t}^{(1,1)}\!=&\pi^2\left(M\!-\!1\right)N\!\left[\left(N\!-\!1\right)M+2\Delta f\!\sum\nolimits_{l=0}^{M-1}\!\!g(l)\right]\!\!\\
\hspace {-0.5pc}R_{p,r,t}^{(2,0)}\!=&\frac{\left(2\pi\Delta f\right)^2\!\left(M\!-\!1\right)MN\left(2M\!-\!1\right)}{6}\\
\hspace {-0.5pc}R_{p,r,t}^{(0,2)}\!=&\frac{\left(2\pi T\right)^2\!\left(N\!-\!1\right)MN\left(2N\!-\!1\right)}{6}+\left(2\pi\right)^2\!N\!\sum\nolimits_{l=0}^{M\!-\!1}\!\!\!g^2(l)\nonumber\\&+\left(2\pi\right)^2T\left(N\!-\!1\right)N\!\sum\nolimits_{l=0}^{M\!-\!1}\!\!g(l)
\vspace{-1mm}
\end{align}
\end{subequations}
\endgroup

\begingroup\makeatletter\def\f@size{9.5}\check@mathfonts
\noindent Further, using the identity $\left(\mathbf{a}\odot \mathbf{b}\right)^{\dagger}\!\left(\mathbf{c}\odot \mathbf{d}\right)\!=\!\left(\mathbf{a}\odot \mathbf{d}\right)^{\dagger}\!\left(\mathbf{c}\odot \mathbf{b}\right)$, and substituting {\small$\mathbf{V}_{p}\approx\eta_{pt}\hat{\mathbf{h}}_{pt}^{\phantom{\dagger}}\hat{\mathbf{h}}_{pt}^{\dagger}$} yield
\endgroup
\begingroup\makeatletter\def\f@size{9.5}\check@mathfonts
\begin{equation}
\begin{aligned}
\hspace {-0.5pc}&\mathbf{h}_{tr}^{\dagger}\mathbf{h}_{tr}^{\phantom{\dagger}}\!=1,\qquad\qquad\qquad\quad\ \ \ \;\!
\mathbf{h}_{pt}^{\dagger}\mathbf{V}_{p}^{\phantom{\dagger}}\mathbf{h}_{pt}^{\phantom{\dagger}}\approx\eta_{pt},
\\
\hspace {-0.5pc}&\dot{\mathbf{h}}_{tr}^{\dagger}\mathbf{h}_{tr}^{\phantom{\dagger}}\!=\mathbf{h}_{tr}^{\dagger}\dot{\mathbf{h}}_{tr}^{\phantom{\dagger}}\!=\frac{M_{\mathrm t}-1}{2},\quad\ \ \,\,\!\dot{\mathbf{h}}_{pt}^{\dagger}\mathbf{V}_{p}^{\phantom{\dagger}}\mathbf{h}_{pt}^{\phantom{\dagger}}\approx\eta_{pt}\frac{M_{\mathrm t}-1}{2},
\\
\hspace {-0.5pc}&\dot{\mathbf{h}}_{tr}^{\dagger}\dot{\mathbf{h}}_{tr}^{\phantom{\dagger}}\!=\frac{\left(M_{\mathrm t}-1\right)\!\left(2M_{\mathrm t}-1\right)}{6},\quad\!
\dot{\mathbf{h}}_{pt}^{\dagger}\mathbf{V}_{p}^{\phantom{\dagger}}\dot{\mathbf{h}}_{pt}^{\phantom{\dagger}}\approx\eta_{pt}\Big(\!\frac{M_{\mathrm t}-1}{2}\!\Big)^2\!\!.\!\!
\label{eq:h_derived}
\end{aligned}
\end{equation}
\endgroup

\noindent Then, by substituting~\eqref{eq:R_derived} and~\eqref{eq:h_derived} into~\eqref{eq:FIM_example}, the equivalent FIM matrix in~\eqref{eq:EFIM} is obtained as
\begingroup\makeatletter\def\f@size{9.5}\check@mathfonts
\begin{equation}
\mathbf{F}_{\boldsymbol{\theta}_{p,r,t}^{(1)}}^{\mathrm{e}}=\frac{2|\beta_{p}|^{2}\eta_{pt}}{\sigma_w^2}\operatorname{diag}\{d_{11},d_{22},d_{33},d_{44}\},
\label{eq:EFIM_derived}
\end{equation}
\endgroup

\noindent where
\begingroup\makeatletter\def\f@size{9}\check@mathfonts
\begin{align*}
\hspace {-0.5pc}d_{11}=&\frac{\left(M_{\mathrm t}-1\right)\left(2M_{\mathrm t}-1\right)MN}{6} - \frac{\left(M_{\mathrm t}-1\right)^2MN}{4}, d_{22}=0, \\
\hspace {-0.5pc}d_{33}=&R_{p,r,t}^{(2,0)}+\big(R_{p,r,t}^{(1,0)}\big)^2/MN, d_{44}=R_{p,r,t}^{(0,2)}+\big(R_{p,r,t}^{(0,1)}\big)^2/MN.\!\!\!
\label{eq:d_derived}
\end{align*}
\endgroup

Finally, by substituting~\eqref{eq:EFIM_derived} into~\eqref{eq:positionFIM}, the desired result in~\eqref{eq:approxPositionFIM} is obtained following a series of algebraic manipulations.

}

\bibliographystyle{IEEEtran}
\bibliography{IEEEabrv,bibRef}

\begin{thebibliography}{10}
\providecommand{\url}[1]{#1}
\csname url@samestyle\endcsname
\providecommand{\newblock}{\relax}
\providecommand{\bibinfo}[2]{#2}
\providecommand{\BIBentrySTDinterwordspacing}{\spaceskip=0pt\relax}
\providecommand{\BIBentryALTinterwordstretchfactor}{4}
\providecommand{\BIBentryALTinterwordspacing}{\spaceskip=\fontdimen2\font plus
\BIBentryALTinterwordstretchfactor\fontdimen3\font minus
  \fontdimen4\font\relax}
\providecommand{\BIBforeignlanguage}[2]{{%
\expandafter\ifx\csname l@#1\endcsname\relax
\typeout{** WARNING: IEEEtran.bst: No hyphenation pattern has been}%
\typeout{** loaded for the language `#1'. Using the pattern for}%
\typeout{** the default language instead.}%
\else
\language=\csname l@#1\endcsname
\fi
#2}}
\providecommand{\BIBdecl}{\relax}
\BIBdecl

\bibitem{Lu2024Integrated}
S.~Lu \emph{et~al.}, ``{Integrated sensing and communications: Recent advances
  and ten open challenges},'' \emph{IEEE Internet Things J.}, vol.~11, no.~11,
  pp. 19\,094--19\,120, Jun. 2024.

\bibitem{liu2022integrated}
F.~Liu \emph{et~al.}, ``{Integrated sensing and communications: Toward
  dual-functional wireless networks for 6G and beyond},'' \emph{IEEE J. Sel.
  Areas Commun.}, vol.~40, no.~6, pp. 1728--1767, Jun. 2022.

\bibitem{zhang2021enabling}
J.~A. Zhang \emph{et~al.}, ``{Enabling joint communication and radar sensing in
  mobile networks—A survey},'' \emph{IEEE Commun. Surv. Tutor.}, vol.~24,
  no.~1, pp. 306--345, 1st Quart. 2022.

\bibitem{Dehkordi2023Beam}
S.~K. Dehkordi, L.~Gaudio, M.~Kobayashi, G.~Caire, and G.~Colavolpe,
  ``{Beam-space MIMO radar for joint communication and sensing with OTFS
  modulation},'' \emph{IEEE Trans. Wireless Commun.}, vol.~22, no.~10, pp.
  6737--6749, Oct. 2023.

\bibitem{Liu2024Cooperative}
S.~Liu, R.~Liu, Z.~Lu, M.~Li, and Q.~Liu, ``{Cooperative cell-free ISAC
  networks: Joint BS mode selection and beamforming design},'' in \emph{Proc.
  IEEE Wireless Commun. and Netw. Conf. (WCNC)}, Apr. 2024, pp. 1--6.

\bibitem{elfiatoure2023cell}
M.~Elfiatoure, M.~Mohammadi, H.~Q. Ngo, and M.~Matthaiou, ``{Cell-Free massive
  MIMO for ISAC: Access point operation mode selection and power control},'' in
  \emph{Proc. IEEE Global Commun. Conf. Workshops (GC Wkshps)}, Dec. 2023, pp.
  104--109.

\bibitem{behdad2024multi}
Z.~Behdad, {\"O}.~T. Demir, K.~W. Sung, E.~Bj{\"o}rnson, and C.~Cavdar,
  ``{Multi-static target detection and power allocation for integrated sensing
  and communication in cell-free massive MIMO},'' \emph{IEEE Trans. Wireless
  Commun.}, vol.~23, no.~9, pp. 11\,580--11\,596, Sep. 2024.

\bibitem{ammar2021user}
H.~A. Ammar, R.~Adve, S.~Shahbazpanahi, G.~Boudreau, and K.~V. Srinivas,
  ``{User-centric cell-free massive MIMO networks: A survey of opportunities,
  challenges and solutions},'' \emph{IEEE Commun. Surv. Tutor.}, vol.~24,
  no.~1, pp. 611--652, 1st Quart. 2022.

\bibitem{Gong2023Simultaneous}
Z.~Gong, F.~Jiang, C.~Li, and X.~Shen, ``{Simultaneous localization and
  communications with massive MIMO-OTFS},'' \emph{IEEE J. Sel. Areas Commun.},
  vol.~41, no.~12, pp. 3908--3924, Dec. 2023.

\bibitem{Fan2024Power}
Y.~Fan, S.~Wu, X.~Bi, and G.~Li, ``{Power allocation for cell-free massive MIMO
  ISAC systems with OTFS signal},'' \emph{IEEE Internet Things J.}, vol.~12,
  no.~7, pp. 9314--9331, Apr. 2025.

\bibitem{Singh2025Target}
S.~Singh, A.~Nakkeeran, P.~Singh, E.~Sharma, and J.~Bapat, ``{Target detection
  for OTFS-aided cell-free MIMO ISAC system},'' \emph{IEEE Trans. Veh.
  Technol.}, early access, Mar. 11, 2025, doi: 10.1109/TVT.2025.3550135.

\bibitem{Pucci2025Cooperative}
L.~Pucci, T.~Bacchielli, and A.~Giorgetti, ``{Cooperative maximum likelihood
  target position estimation for MIMO-ISAC networks},'' \emph{IEEE Wireless
  Commun. Lett.}, 2025, early access, Mar. 5, 2025, doi:
  10.1109/LWC.2025.3548446.

\bibitem{AbuShaban2018Error}
Z.~Abu-Shaban, X.~Zhou, T.~Abhayapala, G.~Seco-Granados, and H.~Wymeersch,
  ``{Error bounds for uplink and downlink 3D localization in 5G millimeter wave
  systems},'' \emph{IEEE Trans. Wireless Commun.}, vol.~17, no.~8, pp.
  4939--4954, Aug. 2018.

\bibitem{shen2018fractional}
K.~Shen and W.~Yu, ``{Fractional programming for communication systems—Part
  I: Power control and beamforming},'' \emph{IEEE Trans. Signal Process.},
  vol.~66, no.~10, pp. 2616--2630, May 2018.

\end{thebibliography}

\end{document}